\documentstyle[11pt,paspiau199,twoside,epsf]{article}

\begin{document}

\title{Associated H$\,${\sc i}\ in Absorbers at High Redshift}

\author{R.C. Vermeulen}

\affil{Netherlands Foundation for Research in Astronomy,\newline P.O.
  Box 2, NL--7990 AA, Dwingeloo, The Netherlands}

\begin{abstract}
  WSRT observations have provided a first inventory of the incidence of
  H$\,${\sc i}\ 21~cm line absorption associated with AGN at redshifts
  up to z=1.0. There is a large range in line depths, from $\tau=0.44$
  to $\tau\le0.001$, and a substantial variety of line profiles, from
  Gaussians of less than ten km~s$^{-1}$ to more typically a few
  hundred km~s$^{-1}$, as well as irregular and multi-peaked
  absorption, sometimes spanning many hundreds of \hbox{km~s$^{-1}$}.
  The chance to detect appreciable H$\,${\sc i}\ absorption is greatest
  in the most compact radio sources, GPSs and CSOs, where it can occur
  in circumnuclear ``disks'' or ``tori'', as well as in gas enveloping
  jets and hot spots; inferred densities range at least between 10
  cm$^{-3}$ and 10$^{4}$ cm$^{-3}$. But H$\,${\sc i}\ absorption occurs
  also in some CSSs, perhaps associated with jet-cloud interaction
  regions, and in quasars with a large optical reddening. VLBI
  observations at the unusual UHF frequencies of redshifted H$\,${\sc
    i}\ 21~cm can give unique ``sight''lines into the physics and
  evolution of young radio sources and their inner galactic medium.
\end{abstract}

\section{Introduction}

Until recently, the galactic medium at cosmological redshifts was
largely unexplored, but after the advent at the WSRT of wideband UHF
receivers, spanning 700--1200 MHz, systematic surveys for H$\,${\sc i}\ 
21~cm line absorption associated with radio-loud active galaxies in the
redshift range $z=0.2$ to $z=1.0$ have been done by different
collaborations and with a variety of goals. The emphasis has been on
bright, compact radio sources, which were a priori thought to be likely
targets for successful detections, with the potential to use the lines
to draw astrophysically interesting conclusions. Indeed, in the early
H$\,${\sc i}\ 21~cm absorption line search of galaxies at relatively
low redshifts by van Gorkom et al.\ (1989), the 4 detections (out of 29
objects observed) were all in CSOs or sources with similarly compact
structure. Absorption was detected at the WSRT in 27/84 targets
analysed to date; highlights and trends will be discussed here.

\section{CSOs}

Compact Symmetric Objects (CSOs) are defined as having radio lobes
visible on both sides of the central engine but not extending over more
than 1~kpc
%(i.e.\ within the narrow line emitting region of the active galaxy)
(Wilkinson et al.\ 1994).
%CSOs at $z<1$ tend to be associated with galaxies, although there are
%counter-examples.
Many CSOs have an overall Gigahertz-Peaked Spectrum (GPS); others are
amongst the smallest Compact Steep-Spectrum souces (CSSs).
%, with somewhat lower frequency turnovers.
The CSS class also contains
%their slightly more extended cousins, termed 
Medium-size Symmetric Objects
(MSOs, Fanti et al.\ 1995), ranging in size up to 15 kpc.
%(still within the confines of the host galaxy), and with the radio
%emission again dominated by a pair of lobes.
A recent review was given by O'Dea~(1998).

It is now commonly thought that CSOs and MSOs are the early stages of
an evolutionary sequence ending up at powerful extended radio sources.
Back-extrapolated hotspot advance speeds imply total radio source ages
of less than 1000 years in several CSOs (Owsianik \&\ Conway 1998;
Owsianik, Conway, \&\ Polatidis 1998; Owsianik, Conway, \&\ Polatidis
1999). MSOs are thought to have ages ranging between 10$^4$ and 10$^5$
years (Fanti et al.\ 1995). However, while the enveloping media
probably do not permanently ``frustrate'' the radio sources in their
growth, jet-cloud or more generally jet-envelope interactions in the
narrow line region of the inner galaxy could well be dynamically
important.
%It is by no means certain that hotspot advance occurs
%steadily, rather than in leaps-and-bounds like a ``dentists drill''
%(Scheuer 1970\check), as the jets interact with their enveloping
%medium. Furthermore, in order to explain the relative numbers of CSOs,
%MSOs, and classical FR-II sources, the luminosities must decrease by
%factors of 10--30 as the sources age (e.g., Snellen et al.\ 1998
%\check). This suggests models in which the expanding radio lobes evolve
%in ram pressure equilibrium with a surrounding medium which decreases
%in density with distance from the AGN (e.g., Snellen or Begelman 1998
%\check).

With the WSRT, 6/7 CSOs in the complete PR sample (Pearson \&\ Readhead
1988) were surveyed for H$\,${\sc i}\ absorption (0710+439 had
excessive external interference). This yielded 4/6 detections
(Vermeulen et al.\ in preparation); peak optical depths and limits are:
0108+388, 44\%; 0404+768, 2.5\%; 1031+567, $<$1\%; 1358+624, 0.4\%;
2021+614, $<$0.2\%; 2352+495, 1.5\% (shown in Figure~1 as a
representative example). Some of the line profiles can be reasonably
approximated by a single Gaussian, but others have a rather more
complex structure; the velocity widths range from 10 \hbox{km~s$^{-1}$}
to 300 \hbox{km~s$^{-1}$}. In fainter CSOs similar absorption lines
have been detected for 5/10 sources (Tschager et al.\ in preparation)
in the CJF sample (Taylor et al.\ 1996a) and 1/2 additional sources in
the COINS sample (Peck et al.\ 2000), but a comparison with the
brighter PR CSOs is not yet meaningful since detection limits $\le$1\%\ 
were not always reached.

\begin{figure}[t] 
\plotfiddle{vermeulen_image.ps}   {3.5cm}{0}{34}{34}{-195}{-131}
\plotfiddle{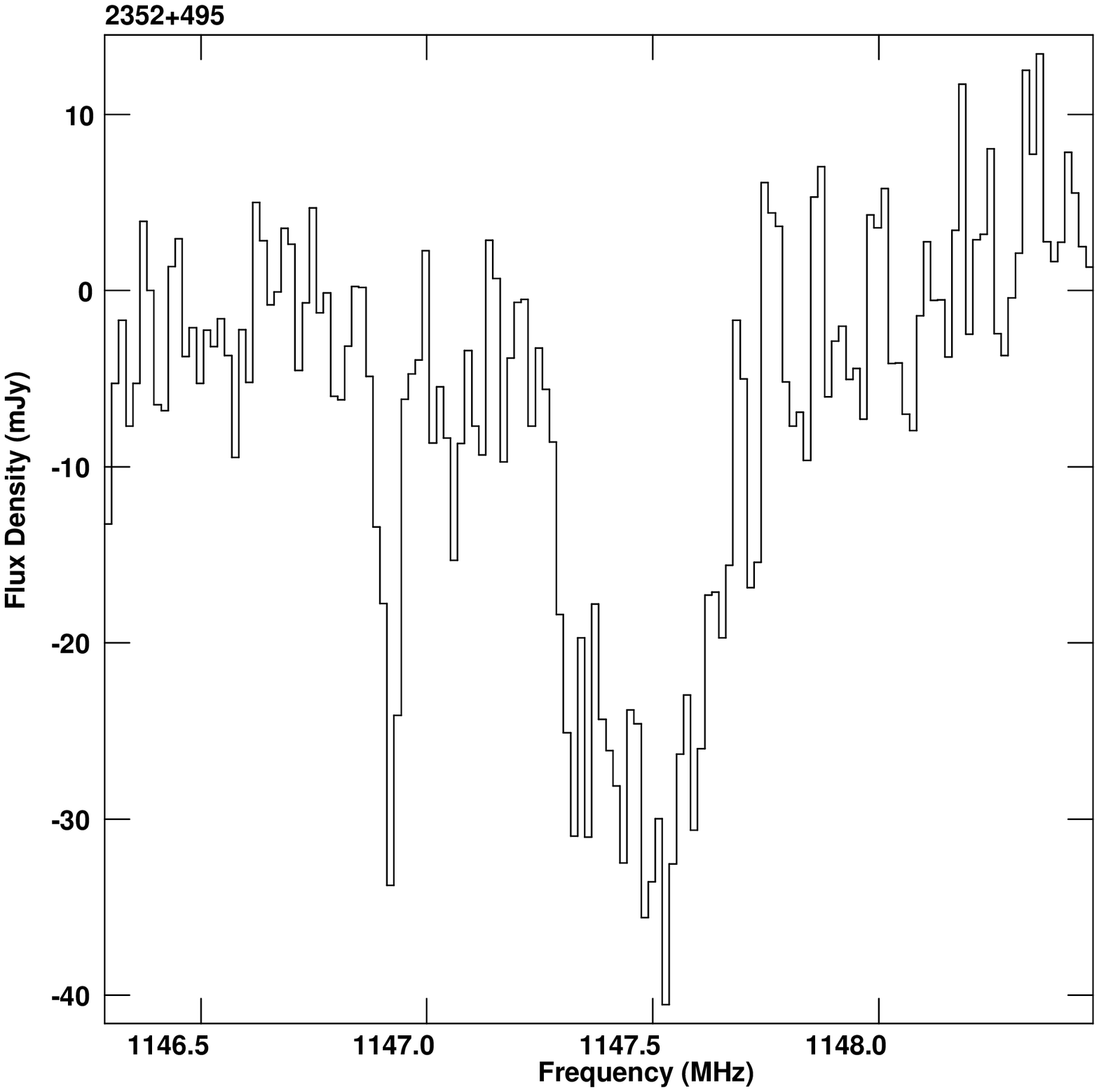}{3.5cm}{0}{33}{33}{0}{-5.5}
\caption{Observations at 1147 MHz (H$\,${\sc i}\ 
  at $z=0.24$) of the CSO 2352+495. Left: continuum VLBI image. Right:
  Spectrum of the central component, which shows absorption.}
\end{figure}

Preliminary analysis of a VLBI observation at 1147 MHz (H$\,${\sc i}\ 
at $z=0.24$) suggests that the absorption in 2352+495 may be confined
to the central part of the continuum structure (shown in Figure~1),
$\le$50 pc across (Vermeulen et al.\ in preparation). The absorption
line profile is then likely to be probing a toroidal or disk-like
region around the active nucleus, and perpendicular to the inner jets.
Disks or tori with a thickness of only a few pc have been demonstrated
to cover a portion of the receding jet in the lower redshift galaxies
Cygnus A (Conway \&\ Blanco 1995, Conway 1999) and NGC 4261 (van
Langevelde et al.\ 1999).  But in the radio galaxy Hydra A (Taylor
1996), and in the low-redshift CSO 4C31.04 (Conway 1996), a
geometrically thick disk, extending over several tens of pc, covers
part of the approaching jet as well. It seems that a thick disk/torus
scenario might apply to 2352+495: the central radio component in
Figure~1 is probably almost entirely from the approaching (Doppler
boosted~?) jet, because the true core is not visible due to synchrotron
self-absorption, free-free absorption, or both (see Taylor, Readhead,
\&\ Pearson 1996b).

Maloney (1999) has argued that circumnuclear tori or disks could be in
a largely atomic state at $T\sim8000$~K; adopting this as the spin
temperature yields a column density of $N({\rm H})\ga10^{21}$~cm$^{-2}$
for the narrowest absorption feature in 2352+495, and $N({\rm
  H})\sim10^{24}$~cm$^{-2}$ for the deep absorption in 0108+388.
Subject to substantial uncertainty in the actual size and geometry of
the covering region, the absorbers have particle densities of $n({\rm
  H})\ge10^2$~cm$^{-3}$ for the narrow absorber in 2352+495 to $n({\rm
  H})\ge10^4$~cm$^{-3}$ for 0108+388.

However, in another PR CSO, 0404+768, preliminary VLBI observations at
886 MHz (H$\,${\sc i}\ at $z=0.60$) indicate that the total flux
density of the central region is insufficient for
%to explain the flux density in 
the main absorption line. It probably arises against part or
all of the outer lobes, which together extend over almost 1 kpc
(Vermeulen et al.\ in preparation). The inferred atomic gas density is
$n({\rm H})\ge10$~cm$^{-3}$ (limited by assuming uniform coverage).
Detailed VLBI observations may be able to probe in a cosmologically
distant galaxy the properties of an atomic gas disk, which could be
analoguous to those found through H$\,${\sc i}\ absorption in nearby
Seyfert galaxies by Gallimore et al.\ (1999). But in addition, VLBI
spectral imaging will also reveal whether and how the radio jets and
lobes in young powerful radio sources interact with their gaseous
environment (for which Seyfert galaxies like IC 5063 might be
low-redshift analogues; see Oosterloo et al.\ 2000).

\section{CSSs}

The recent WSRT suveys show a definite but by no means exclusive trend
of diminishing H$\,${\sc i}\ absorption with increasing linear extent
of the radio sources (Pihlstr\"om et al.\ in preparation); this was
already suspected by Conway (1996). The larger sources are well outside
the torus and accretion disk area, and probably probe regions with a
lower atomic gas column density.

The shallowest absorption line detected to date, in the CSS galaxy
3C213.1, has only 0.1\%\ peak depth, but in 3C49 (3.6\%) and 3C268.3
(12\%) prominent absorption exists (de Vries et al.\ in preparation).
In both of these, (part of) one of the radio lobes is coincident, at
least in projection, with an optically prominent feature or knot in the
host galaxy. If VLBI observations confirm that the optical knots are
the sites of the H$\,${\sc i}\ absorption, this offers unique
opportunities to study the ``alignment effect'' and details of
jet-cloud interactions in cosmologically distant sources. A jet-cloud
interaction site is also the most plausible location of the H$\,${\sc
  i}\ absorption found in the ``superluminal CSS'' quasar 3C216, where,
because of its orientation, the circumnuclear disk or torus is unlikely
to be visible in absorption (Pihlstr\"om et al.\ 1999).

\section{Quasars}

Normal core-dominated quasars are unlikely to show H$\,${\sc i}\ 
absorption (Pihlstr\"om et al.\ in preparation), but 2/3 non-CSS
quasars pre-selected to have high optical reddening (from Stickel et
al.\ 1996) were found to have significant H$\,${\sc i}\ absorption
(probably associated, intervening systems are not dealt with here):
0500+019, 3.6\%; 1504+377, 34\%; 2149+056, $<$1.5\% (Carilli et al.\ 
1998).

\smallskip
\acknowledgements

The WSRT H$\,${\sc i}\ surveys of CSOs and CSSs summarised here were
collaborative projects with P.D. Barthel, S.A.  Baum, R.  Braun, J.E.
Conway, W.H. de Vries, C.P. O'Dea, Y.  Pihlstr\"om, H.J.A.
R\"ottgering, R.T. Schilizzi, I.A.G. Snellen, G.B.  Taylor, and W.
Tschager.

\footnotesize % Necessary in order to stay within 4 page limit !

\end{document}